\documentclass[amscd,amssymb,verbatim]{amsart}
\usepackage{epsfig}
\usepackage{graphics}
\begin{document}
\numberwithin{equation}{section}
 \title{Finite Groups and  Quantum
Yang-Baxter Equation} 
\author{Peter Varga}
\address{Institute of Mathematics and Informatics \\
  Lajos Kossuth University\\
  H-4010 Debrecen, Hungary} \email{varga\@@math.klte.hu} 
\thanks{This research was partially supported by OTKA grant F-015470.}
\date{}

\begin{abstract}
We construct  integrable modifications of 2d lattice gauge theories with finite
gauge groups.  
\end{abstract}

\maketitle

The solvability of many 2d lattice statistical models is closely
connected to the Quantum Yang-Baxter equation (QYBE)
\cite{Y,B}. Solutions of the QYBE are equivalent to weight functions
of vertex models.  

Probably the most simple 2d integrable system is (lattice) gauge
theory. The weights of the field configurations around a plaquette
satisfy the QYBE Fig.1a. (The gauge group is assumed to be finite.)
\begin{equation}\label{E:1}
w(a,b,c,d)=R_{a,b}^{d^{-1},c^{-1}}(\lambda)=
\sum_{r \in R(G)}\lambda^{r}\chi_{r}(abcd)
\end{equation}
\begin{equation}\label{E:2}
\sum_{g,h,i \in G} 
  R_{a,b}^{g^{-1},i}(\lambda)R_{i^{-1},c}^{h,d}(\mu)R_{g,h^{-1}}^{f,e}(\nu)=
\sum_{g,h,i \in G}
R_{b,c}^{h^{-1},g}(\nu)R_{a,h}^{f,i^{-1}}(\mu)R_{i,g^{-1}}^{e,d}(\lambda).
\end{equation}
($a,b,c,\dots $ are elements of the group $G$, $\chi_{r}$ is a
character of some irreducible representation $r \in R(G)$. \cite{M})

\vspace{12pt}
\scalebox{0.6}{
\includegraphics*{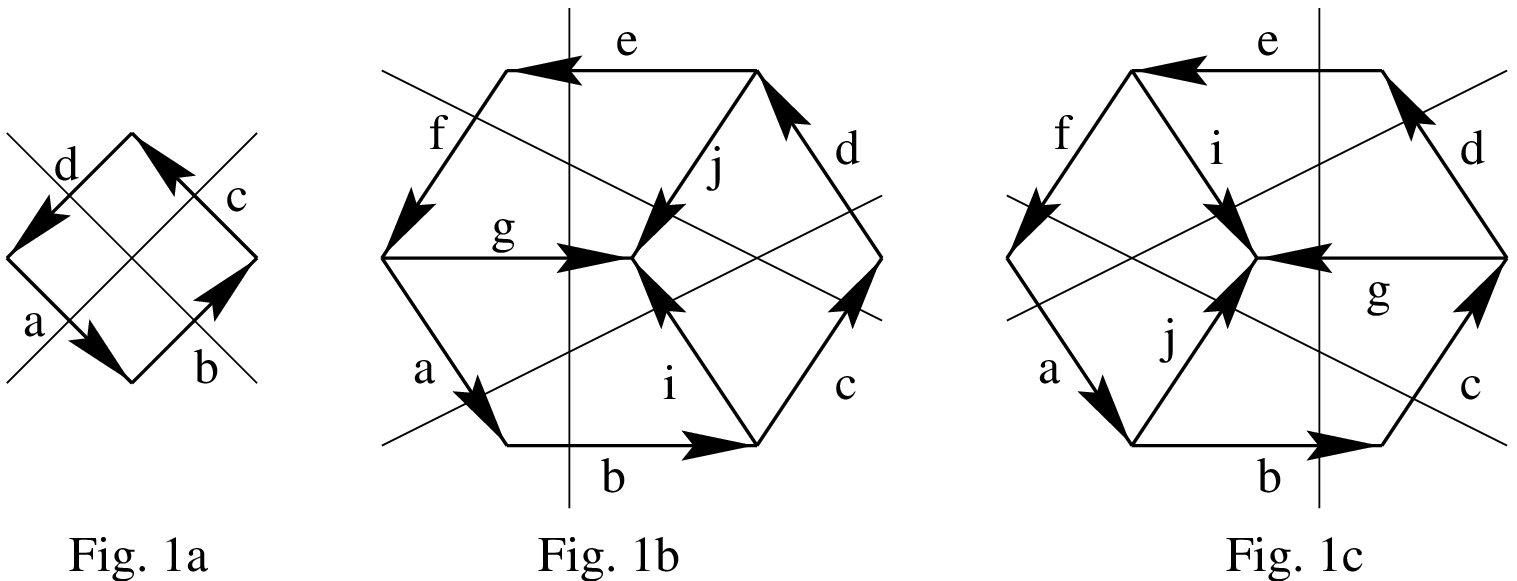}}
\vspace{6pt}

Equation \eqref{E:2} is satisfied since since both side depend on only
the holonomy $abcdef$ around the three plaquettes, which is the same
on Fig.1b and Fig.1c. A more direct proof is based on the
character identity
\begin{equation}\label{E:3}
\sum_{x \in G} \chi_{r}(ax) \chi_{s}(x^{-1}b) =
\frac{|G|}{d_{r}}\delta_{rs} \chi_{r}(ab),
\end{equation}
where $d_{r}$ is the dimension of the representation $r$. Using
\eqref{E:3} three times, both sides of \eqref{E:2} evaluate to
\begin{equation}\label{E:4} 
\sum_{r \in G} \lambda^{r} \mu^{r} \nu^{r} \frac{|G|^{3}}{d_{r}^{2}}
\chi_{r}(abcdef).
\end{equation}

A slight modification  of \eqref{E:1} also satisfies  QYBE:
\begin{align} \label{E:5}
&w(a,b,c,d)=R_{a,b}^{d^{-1},c^{-1}}(\lambda)= \\ \notag
&\sum_{r \in R(G)} \biggl(
\lambda_{0}^{r}\chi_{r}(abcd)+
\lambda_{1}^{r}\bigl(\chi_{r}(bd)+\chi_{r}(ac)\bigr)\biggr).
\end{align}
At this choice of weights $w$ both sides of \eqref{E:2} equal to
\begin{equation}
\begin{aligned}
|G|^{3}\sum_{r\in G}d_{r}^{-2}
\biggl\{
&\lambda_{0}^{r}\mu_{0}^{r}\nu_{0}^{r}\chi_{r}(abcdef)+
\lambda_{0}^{r}\mu_{1}^{r}\nu{1}^{r}\chi_{r}(abde)\\
+&
\mu_{0}^{r}\nu_{1}^{r}\lambda_{1}^{r}\chi_{r}(cdfa)
+
\nu_{0}^{r}\lambda_{1}^{r}\mu_{1}^{r}\chi_{r}(efbc).
\biggr\}
\end{aligned}
\end{equation}

If $G$ is a compact Lie-group, \eqref{E:5} can be obtained as the
weights of the lattice version of the continuous 2d Lie-algebra valued
vector field model with action
\begin{equation}\label{E:6}
S=\sum_{a=1}^{\dim{G}}\biggl\{
\frac{1}{2}{F_{xt}^a}^{2}+\alpha\bigl[
(\partial_{x}A_{t}^{a})^{2}+(\partial_{t}A_{x}^{a})^{2}\bigr]
\biggr\},
\end{equation}
where $A_{x}^{a},A_{t}^{a}$ are the spatial end the temporal
components of the vector field. (We assumed that the Killing-metric on
$G$ is $\delta_{a,b}$.) Since \eqref{E:6} is a continuous version of an
integrable quantum system, it is reasonable to believe that the
classical Euler-Lagrange equation are also integrable ones. The
transfer matrices of the lattice system commute for different $\alpha$
and $\beta$ parameters. However, the classical system is constrained
since $A_{t}$ is not dynamical. The allowable initial conditions live
on different constraint surfaces for different parameters, so it is not
quite clear to us that the integrability of the lattice version really
implies the integrability of the classical version.

After this short digression we return to the lattice world and present
another modification of lattice gauge theory. The weigh of a field
configuration around a plaquette is given by
\begin{equation}\label{E:7}
w(a,b,c,d)=R_{a,b}^{d,c}=
\sum_{\{\sigma_{x}=\pm1\}}
\sum_{r \in R(G)} \lambda^{r}\chi_{r}
(a^{\sigma_{a}}b^{\sigma_{b}}c^{\sigma_{c}}d^{\sigma_{d}}).
\end{equation}
As it does not matter if the variable assigned to a link is $g$ or
$g^{-1}$ the set of link variables are the equivalence classes $\tilde
G= G/\{g \sim g^{-1}\}$. For this weight system QYBE does not hold
automatically. The summation over the variables $g,h,i$ generate terms
like (on Figure 1b):
\begin{equation}\label{E:8}
\sum_{g \in G} \chi_{r}(a^{\sigma_{a}}b^{\sigma_{b}}i^{\sigma_{i}}g)
\chi_{s}(g h^{\sigma_{h}}e^{\sigma_{e}}f^{\sigma_{f}})=
\frac{|G|}{d_{r}}\delta_{r\bar{s}}
\chi_{r}(a^{\sigma_{a}}b^{\sigma_{b}}c^{\sigma_{c}}
       f^{-\sigma_{f}}e^{-\sigma_{e}}h^{-\sigma_{h}}).
\end{equation}
($\bar{s}$ is the complex conjugate of the representation $s$.) Since
the cyclic order of the variables $a,b,f,e$  would change a different way
on Fig.1c, the QYBE is not necessarily satisfied. There are two
ways to avoid this problem. The first is to require that if
$\lambda_{r}\neq 0$ then $\lambda_{\bar{r}}=0$. Unfortunately, in this
case the weights of the variable configurations are not real, so they
are not the weights of a statistical mechanical system. The second
method is to use abelian groups, so the order of the group elements is
irrelevant. 

At last we investigate the ground state structure of these models.  
Two dimensional discrete lattice gauge theory has infinitely
degenerate ground state structure, which prevents the occurrence of of
phase transitions. The ground state degeneracy is somewhat lifted in
the model \eqref{E:5}, however, the number of ground state
configurations is still infinite, since the configuration on
Fig.2 has the same weight as the configuration where all the link
variables are equal to $e$ if $a_{i}$ and $b_{i}$ are in an abelian
subgroup  $G_{A}\subset{}G$.

\vspace{18pt}
\scalebox{0.7}{
\includegraphics*{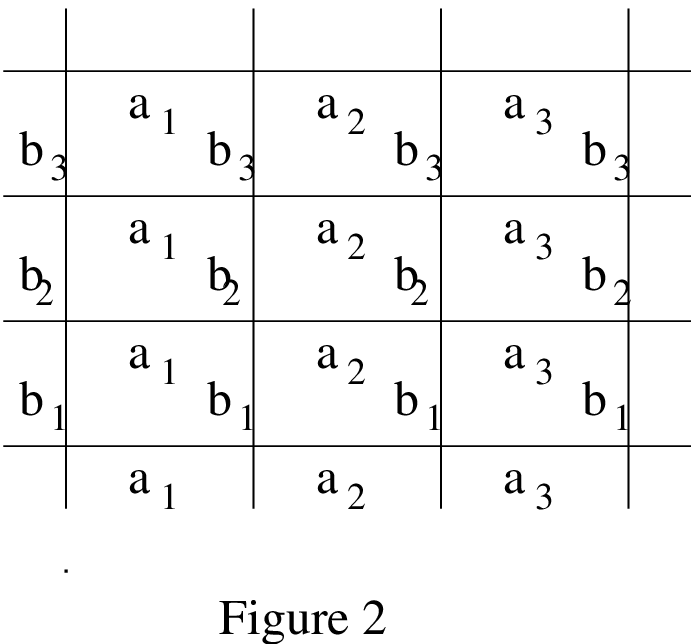}}
\vspace{8pt}

In  contrast to the previous cases, a system with weights \eqref{E:7}
has unique ground state if and only if there is no involution
$p=p^{-1}$ in $G$, where we assume that $G$ is abelian and
$\tilde{w}(g)=\sum_{r} \lambda{}^{r}\chi{}_{r}(g)$ is minimal at $g=e$.
   If there is
 an involution $p\in G$, then the
configurations where $a_{i}$ and $b_{i}$ are either $e$ or $p$ have
the same weight as the configuration where all the link variables are
equal to $e$. The absence of involutions in an abelian group implies
to that $|G|$ is odd.
If $\tilde{w}(g)$ has minimum at some $g\neq{}e$, then the ground
state is infinitely degenerate. Indeed, let us mark a subset of the
links so that each plaquette has exactly one marked side, and set the
marked link variables to $g$ (or $g^{-1}$) and set to $e$ the rest of the
links. Such configuration is  ground state. Since the marking of the
links can be done in many ways, the ground state is highly degenerate.
If $\tilde{w}(g)$ has two minimums at $g_{1}$ and $g_{2}$, then the
marked links can be set either to  $g_{1}$ or  $g_{2}$ in a completely
 random manner, so the Gibbs-state remains unique even in this case.
 Consequently the Phase structure is trivial.

\end{document}